\documentclass[11pt]{article}
\usepackage[T1]{fontenc}
\usepackage[utf8]{inputenc}
\usepackage{lmodern}
\usepackage[margin=1in]{geometry}
\usepackage{graphicx}
\usepackage{booktabs}
\usepackage{adjustbox}
\usepackage{caption}
\usepackage{amsmath}
\usepackage{listings}
\usepackage[hidelinks]{hyperref}
\usepackage{url}
\title{Completeness and Cost of Retrieval Structures for Multi-Tool Requests in LLM Agents}
\author{Shuyang Zhang\\
\small The Hong Kong Polytechnic University\\
\small\texttt{25049993g@connect.polyu.hk}}
\date{}
\hypersetup{pdftitle={Completeness and Cost of Retrieval Structures for Multi-Tool Requests in LLM Agents},pdfauthor={Shuyang Zhang}}
\begin{document}
\maketitle
\begin{abstract}
An agent that must call several tools to fulfill a request fails when the tool retriever omits any one of them, yet tool retrievers are mostly built and evaluated on queries requiring a single tool. We compare representatives of six retrieval structures (flat lexical and dense retrieval, LLM query decomposition, LLM-navigated hierarchies, server-then-tool matching, dependency graphs, and trained set-level retrievers) across two multi-tool testbeds: 345 AppWorld queries with expert dependency labels and 103 MCP-Bench tasks over 257 MCP tools. We measure completeness@k, the share of queries whose entire required tool set appears in the top k, alongside the LLM calls, tokens, and latency that each structure incurs at retrieval time. A motivating analysis on all 7,961 ToolRet queries shows that aggregate completeness mixes two regimes: completeness at 10 is .58 to .67 on single-target queries and .08 to .17 on queries with four or more targets, and the ordering of retrievers on single-target queries reverses on multi-target ones. Across the two testbeds, none of the compared structures that is untrained on the testbed and makes at most one LLM call per query completes more than 10\% of AppWorld queries at k = 16. The two structures that do better incur different costs: a set-level retriever trained on 75 to 90 in-domain queries reaches .33 on AppWorld and .58 on MCP-Bench with no LLM calls, while an LLM-navigated hierarchy reaches .30 and .40 with a median of 31 to 38 LLM calls and about five minutes per query. Their failure modes also differ: the trained retriever mostly ranks the missing tools just below the cutoff, whereas the hierarchy finds the target API and stops before collecting its prerequisites. In a downstream check with the official AppWorld agent, tasks with a complete retrieved set succeeded 21 of 21 times, whereas incomplete ones succeeded 2 of 39 times.

\end{abstract}

\section{Introduction}
Tool-using language-model agents increasingly work with tool catalogs that are too large, or too costly, to place in the prompt at every step, so a retriever decides which tools the agent sees \cite{r1,r2,r3}. Many requests need several tools, often because one call produces an identifier or credential that another call consumes \cite{r4,r5}. For such a request, the retriever has to return the whole required set: if one required tool is missing, the agent cannot carry out that step. This is why COLT framed tool retrieval in terms of completeness \cite{r6} and why a recent evaluation of deployed tool search reports completeness alongside recall \cite{r7}. We ask which retrieval structures return complete tool sets for multi-tool requests, what each spends at retrieval time to do so, and how each fails when it does not.

Existing evaluations do not answer this question, for two reasons. First, tool-retrieval benchmarks are dominated by single-tool queries and report aggregate scores. On ToolRet \cite{r8}, 55\% of the 7,961 queries have one target tool, and for those queries completeness is simply a hit. Second, the structures proposed for multi-tool retrieval have been evaluated on different data against different baselines. Flat dense retrievers \cite{r8}, LLM query decomposition and planning \cite{r9,r10}, LLM-navigated category hierarchies \cite{r11}, server-then-tool matching \cite{r3}, dependency graphs over tools or skills \cite{r12,r13} and trained set-level retrievers \cite{r6,r14} rest on different assumptions about how a multi-tool need is found, and they spend different resources: labelled training queries, LLM calls at query time, or graph construction. Zhang et al. \cite{r10} compared flat retrieval with an LLM plan-then-retrieve paradigm on ToolBench and found that which paradigm is better depends on whether a query needs one tool or several. It remains open whether this dependence extends to the other families, which of them reach complete sets on requests with dependencies between tools, and at what cost.

We compare representatives of six retrieval structures on two multi-tool testbeds: 345 AppWorld \cite{r5} queries with the expert parameter-level dependency labels of In-N-Out \cite{r4}, each needing at least two APIs, and 103 MCP-Bench \cite{r15} tasks over 257 tools of real MCP servers, 97 of which need several tools. Each method is run as its authors designed it, the trained set-level methods on each testbed's training queries and the others without training, with one embedding model shared where a method allows it and one LLM for every LLM step. We measure completeness@k and recall@k, the LLM calls, tokens and latency each method spends per query, and, for every incomplete query, which required tools were missed: the target the request asks for, a prerequisite whose consumer was retrieved, or a tool ranked just below the cutoff. Before the main comparison, we use all 7,961 ToolRet queries to show how much an aggregate score conceals; after it, we run the official AppWorld agent on 60 tasks to check whether completeness relates to task success.

The study yields four findings.

\begin{enumerate}
\item Aggregate completeness on ToolRet mixes two regimes. Completeness at 10 is .58 to .67 on single-target queries and .08 to .17 on queries with four or more targets, while recall at 10 on multi-target queries stays between .41 and .61. bge-base, best or tied for best on single-target queries, is overtaken on queries with two or three targets by both decomposition and COLT's released retriever. This agrees with Zhang et al. \cite{r10} and extends their observation to completeness, to queries with up to four or more targets and to trained and graph-based retrievers.
\item On AppWorld and MCP-Bench, flat lexical and dense retrieval, query decomposition, server-then-tool matching and dependency-graph retrieval complete at most 9.6\% of AppWorld queries and at most 36\% of MCP-Bench tasks at k = 16. Two structures do better, by spending different resources. COLT, trained on 75 to 90 labelled queries from the target pool, reaches completeness at 16 of .328 and .583 without any LLM call at query time. AnyTool's LLM-navigated hierarchy needs no training and reaches .301 and .398, with a median of 31 and 38 LLM calls and about 300 seconds per query. The two are statistically indistinguishable on AppWorld; COLT leads on MCP-Bench.
\item The structures fail in different ways. On AppWorld, 58\% of the tools COLT misses are ranked between 17 and 50. AnyTool usually retrieves the API that the request names but not the APIs that produce its inputs, which is the primary cause in 162 of its 241 incomplete queries. Flat and graph retrievers and the simplified HYSET mostly miss the target API itself.
\item With the official AppWorld agent, exposing COLT's top 16 APIs led to success on 21 of 21 tasks where that set was complete and on 2 of 39 where it was not, while exposing the gold API set succeeded on 26 of those 39 incomplete tasks.
\end{enumerate}

\section{Related Work}
Tool retrieval benchmarks. ToolRet \cite{r8} collects 7.6 thousand retrieval tasks over about 43 thousand tools from many tool-use datasets and evaluates sparse, dense, re-ranking and tool-specific retrievers; it reports recall and completeness at 10 as aggregates. Zhu et al. \cite{r16} show that single-label annotations underestimate tool retrievers and expand the labels with equivalent tools. LiveMCPBench \cite{r17} and MCP-Bench \cite{r15} evaluate agents over many MCP servers, and GatewayBench \cite{r18} compares deployed MCP gateways on capability and security. Its gateways implement several of the structures we study: flattened gateways expose every tool, and hierarchical ones let the agent find tools by layered discovery or semantic search during execution. GatewayBench measures each gateway as a product, end to end; we hold shared components fixed and measure the retrieved set and its retrieval-time cost. These benchmarks measure agents or individual retrievers; they do not stratify retrieval completeness by the number of required tools or compare structural families on it.

Retrieval structures. Flat retrievers score each tool description against the query, lexically as BM25 does \cite{r19} or by embedding both in one space \cite{r20,r21,r22}; Re-Invoke enriches tool documents with synthetic queries \cite{r23}, and ToolDreamer instills LLM reasoning into tool retrievers \cite{r24}. Decomposition and planning methods split the request into sub-needs before retrieval, trained with reinforcement learning in TOOLQP \cite{r9}, prompted zero-shot in the plan-then-retrieve paradigm of Zhang et al. \cite{r10}, or combined with tool description augmentation \cite{r25}. Hierarchical methods narrow the catalog step by step: AnyTool \cite{r11} uses LLM agents over a category, tool and API hierarchy, MCP-Zero \cite{r3} matches requested server domains before tools, and A2X \cite{r26} builds a service taxonomy with an LLM. Hi-RAG \cite{r27} exploits the type, service and tool structure of MCP. Dependency-graph methods add edges between tools: Tool Graph Retriever \cite{r13} learns pairwise dependencies, and Graph-of-Skills \cite{r12} induces producer-consumer edges from input and output schemas and ranks with personalized PageRank. Set-level methods score tool sets rather than single tools: COLT \cite{r6} adds a collaborative learning stage over query, scene and tool graphs, and HYSET \cite{r14} predicts query-conditioned hyperedges. Our comparison includes a representative of each of these families.

Comparisons of structures. Zhang et al. \cite{r10} compare retrieve-then-plan and plan-then-retrieve on ToolBench and find a crossover between single-tool and multi-tool queries that holds across retrieval budgets, registry sizes and two LLMs; the plan step roughly doubles inference cost. Kolluru and Sportsman \cite{r28} compare a hybrid ranker with a typed skill graph and explain why graph edges drawn from embedding neighbourhoods cannot extend retrieval reach. Partee et al. \cite{r7} evaluate a deployed two-stage tool recommender by completeness at 10 on ToolRet and by billed session cost. Our study differs from these in three respects: it compares six structural families rather than two paradigms or one system, it measures completeness on testbeds with dependency labels, and it attributes each family's failures to missing targets, unrecognized prerequisites or ranking depth.

\section{Motivation: Aggregate Scores Hide the Multi-Tool Regime}
We first ask what an aggregate completeness score on a standard benchmark says about multi-tool requests. We use all 7,961 ToolRet queries \cite{r8} over 44,453 tools in 35 tool pools, as counted in the released data (the paper reports about 7.6 thousand queries and 43 thousand tools), in ToolRet's controlled setting in which each query is ranked only against the tools of its own sub-dataset, with the query text and no instruction. Because this setting differs from ToolRet's full-corpus table, our absolute numbers are not comparable with the published ones. We group queries by their number of distinct target tools: 4,402 have one target, 1,937 two, 1,089 three and 533 four or more. We compare two general embedders (bge-base-en-v1.5 \cite{r20} and e5-base-v2 \cite{r21}), the released first-stage COLT retriever (a Contriever \cite{r22} fine-tuned on ToolBench; the set-level second stage is not released), the decomposition arm described in Section 4.3, and the deterministic variant of Graph-of-Skills.

Figure 1 and Table 1 show completeness at 10 by stratum. On single-target queries, completeness equals hit rate and lies between .583 and .672. It falls to between .255 and .525 with two targets and to between .079 and .169 with four or more. The aggregate score of bge (.507) is the average of .672 on single-target queries and .303 on the 3,559 multi-target queries. Recall at 10 on multi-target queries stays between .533 and .613 for the four embedding-based arms, so most required tools are retrieved but the set is rarely complete.

\begin{figure}[htbp]
\centering
\includegraphics[width=\linewidth]{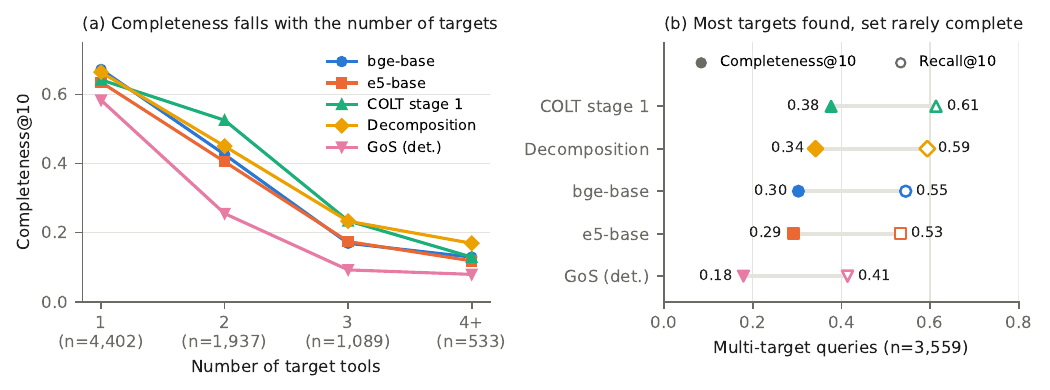}
\caption*{\textbf{Figure 1.} ToolRet, in-sub-dataset retrieval, query only. (a) Completeness at 10 by number of target tools. (b) Multi-target queries (two or more targets, n = 3,559): completeness at 10 (filled) and recall at 10 (open) per retriever; values are query-weighted means of the 2, 3 and 4+ strata.}
\end{figure}

\begin{table}[htbp]
\centering\small
\caption*{\textbf{Table 1.} Completeness at 10 on ToolRet by number of target tools (in-sub-dataset retrieval, query only). Multi = two or more targets.}
\begin{adjustbox}{max width=\linewidth}
\begin{tabular}{lrrrrrr}
\toprule
Retriever & 1 (n=4,402) & 2 (n=1,937) & 3 (n=1,089) & 4+ (n=533) & Multi (n=3,559) & All (n=7,961) \\ \midrule
bge-base & .672 & .426 & .169 & .129 & .303 & .507 \\
e5-base & .635 & .405 & .174 & .118 & .291 & .481 \\
COLT stage 1 & .642 & .525 & .234 & .129 & .377 & .523 \\
Decomposition & .664 & .450 & .233 & .169 & .341 & .520 \\
GoS (deterministic) & .583 & .255 & .092 & .079 & .178 & .402 \\
\bottomrule
\end{tabular}
\end{adjustbox}
\end{table}

The ordering of retrievers changes with the number of targets (paired bootstrap 95\% confidence intervals of the difference in completeness at 10 against bge). On single-target queries, bge is the best or tied for best: COLT stage 1 is lower by between .018 and .041, e5 by between .026 and .047, and decomposition is tied ([-.018, .002]). With two targets, COLT stage 1 is higher than bge ([.077, .122]) and so is decomposition ([.006, .041]); with three, both are higher ([.041, .089] and [.043, .087]); with four or more, decomposition is higher ([.013, .064]) and COLT stage 1 is tied. COLT's released retriever was trained on ToolBench data, so we repeated the analysis without the three ToolBench-derived pools (6,350 queries remain); COLT stage 1 is again lower than bge on single-target queries ([-.047, -.023]) and higher on multi-target ones (.347 vs .317). Scoring at k = 16 gives the same ordering. Tables A3 and A4 give all values. The decomposition arm splits 76\% of single-target queries into more than one sub-request, yet stays tied with bge on them.

These results agree with the crossover that Zhang et al. \cite{r10} report for flat retrieval and LLM planning on ToolBench, and show that it holds for completeness, for queries with up to four or more targets and for a retriever trained on tool data. They also show that a single-target majority makes aggregate completeness a poor guide to multi-tool performance. The rest of the paper therefore evaluates on testbeds in which every AppWorld query and 97 of 103 MCP-Bench tasks need several tools.

\section{Experimental Setup}
\subsection{Task and Metrics}
Given a natural-language request q and a tool pool T, a retriever returns a ranked list of tools. Let R(q) be the set of tools the request requires and L\_k(q) the first k tools of the list. Completeness at k is the share of queries with R(q) contained in L\_k(q); recall at k averages |R(q) $\cap$ L\_k(q)| / |R(q)|. We report k = 4, 8 and 16 on the main testbeds and k = 10 on ToolRet, the value ToolRet reports. Several arms return variable-length sets rather than full rankings; their shorter lists count as returned. We compare arms with paired bootstrap 95\% confidence intervals over queries (2,000 resamples).

Retrieval-time cost is what an arm spends per query after any index or model has been built: the number of LLM calls, the prompt and completion tokens, and the wall-clock latency of the LLM stage. Training data and index construction are reported separately as requirements rather than as per-query cost.

\subsection{Testbeds}
AppWorld. AppWorld \cite{r5} is an executable environment of everyday apps. In-N-Out \cite{r4} provides expert parameter-level edges between AppWorld APIs, and its release includes AppWorld retrieval queries labelled with the APIs each query calls and, for each of their parameters, the APIs that can produce it. We take a query's required set to be these APIs together with their producers. We use the 345 queries of the official test splits (129 test-normal, 216 test-challenge) over the 456 APIs of the In-N-Out pool. AppWorld agents always receive the six supervisor APIs that hold the user's credentials, and AppWorld's ground truth excludes them, so we remove them from rankings and required sets before scoring; app login APIs stay required. The remaining required sets have between 2 and 12 APIs (median 6). From the In-N-Out labels we form the dependency graph of each query's required APIs: 258 queries form a single connected chain (median 4 APIs) and 87 contain several components with at least one dependency (mixed, median 7 APIs); no test query needs a single API or only independent APIs.

MCP-Bench. MCP-Bench \cite{r15} provides tasks over real MCP servers. We list the 257 tools of all 28 servers with MCP-Bench's own collector and use each task's fuzzy description as the query. MCP-Bench has no tool-set labels, so we derive each required set from the tools named in the task's detailed description and dependency analysis, restricted to the task's servers. One task names its tools only by paraphrase and is excluded, leaving 103 tasks (55 on one server, 30 on two, 18 on three; median required set sizes 6, 10 and 11). Six single-server tasks name one tool; the other 97 need several. These sets come from the generator's plans, so they are reference sets rather than unique gold sets.

\subsection{Retrieval Structures and Representatives}
We choose one or two representatives per family; Appendix B lists prompts and all changes to the released code. Among methods that implement a family's core mechanism and have usable code, we prefer formally published ones and use a preprint only when no published method can be run. Shared components are fixed across arms where the method allows: bge-base-en-v1.5 is the embedding model of the flat dense, decomposition, server-then-tool and graph arms, and every LLM step uses DeepSeek-V3.2 at temperature 0.

Figure 2 shows each structure as we run it, with the steps that call an LLM, the components trained on the testbed and the parts built before any query.

\begin{figure}[htbp]
\centering
\includegraphics[width=\linewidth]{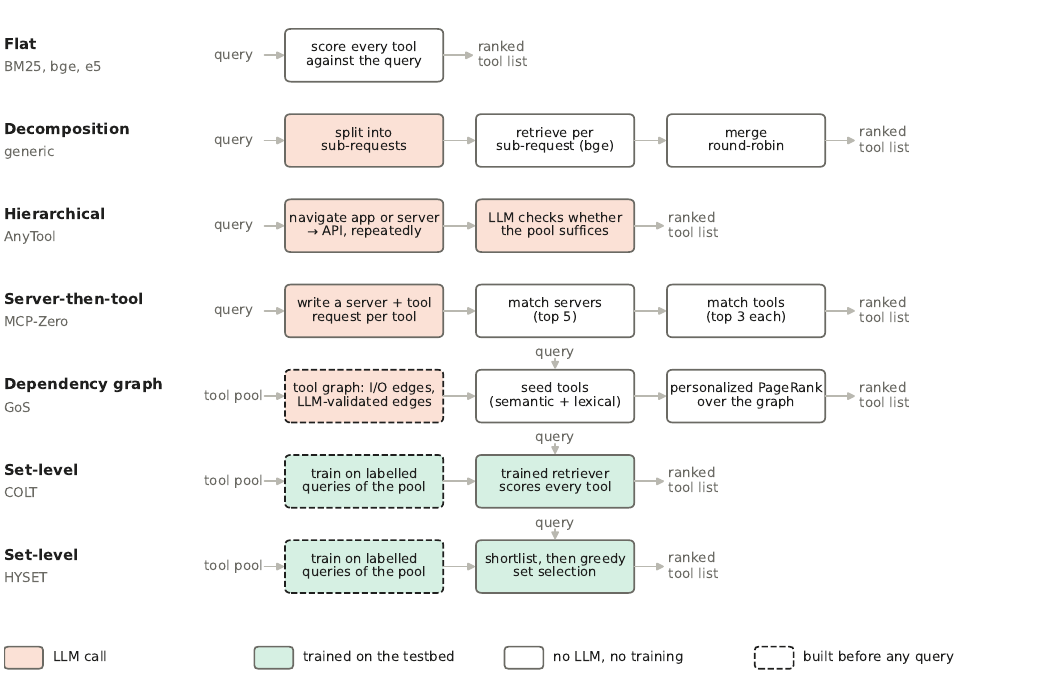}
\caption*{\textbf{Figure 2.} The compared retrieval structures as run in this study. Shaded steps call an LLM (orange) or are trained on the testbed's labelled queries (green); dashed boxes are built once from the tool pool before any query arrives.}
\end{figure}

Flat. Lexical retrieval with BM25 \cite{r19}, and dense retrieval with bge-base-en-v1.5 \cite{r20} and with e5-base-v2 \cite{r21}, over tool descriptions, no training.

Decomposition. No trained decomposition method has public code, so we implement the generic form: one LLM call splits the request into sub-requests, each sub-request retrieves with bge, and the per-sub-request rankings are merged round-robin.

Hierarchical (LLM-navigated). AnyTool \cite{r11} with the official agent code. AppWorld and MCP-Bench have no level above apps and servers, so apps and servers serve as both categories and tools. We run the search stage only and take the API pool at the official stopping rules (at most 64 APIs, 5 leaf tools per agent, a 200 thousand-token cap); the downstream solver and self-reflection are not run. The official function schema for API lists declares a null type, which the LLM obeyed by returning nothing, so we replaced it with an array of objects.

Server-then-tool. MCP-Zero \cite{r3} with its released matcher logic: the LLM writes one request per needed tool naming a server domain and an operation, and the released matcher code keeps the top 5 servers and returns the top 3 tools for each request. The official matcher uses a commercial embedding model; we use bge so that the embedding model is held fixed across arms. Every arm sees the same server text, which is the app or server name only.

Dependency graph. Graph-of-Skills \cite{r12} with its default construction, including LLM-validated semantic edges built at index time, and with the optional LLM query rewrite off, which is its default in the released code, so it makes no LLM call at query time. Its deterministic variant without the LLM is reported as an ablation. On MCP-Bench, no tool declares an output schema, so the deterministic producer-consumer edges do not form (0 edges) and the LLM adds 254 edges; on AppWorld the default build has 4,935 edges.

Set-level (trained). COLT \cite{r6}, a published method with released code, retrained with the authors' configuration (with compatibility changes to the first-stage training loop and the number of negatives) on each testbed because its set-level stage needs labelled queries on the same tool pool. On AppWorld we train on the 90 official training queries and select checkpoints on the 57 development queries; on MCP-Bench we use five-fold cross-validation stratified by the number of servers. Before retraining we reproduced the authors' ToolBench G2 result with their code (R@5 85.26, N@5 82.78, C@5 70.85, against 85.03, 82.54 and 70.10 in the paper). HYSET \cite{r14}, a preprint, is run as a simplified variant trained on the same splits: annotation-only training without the execution reward, whose rollout code is not released, and a maximum set size raised to the largest required set (13 on AppWorld, 20 on MCP-Bench).

Each arm is used as its authors designed it. COLT and HYSET are supervised methods whose published protocols train on labelled queries over the tool pool they retrieve from, so we train them on each testbed's training queries; the other arms are training-free by design and are run without training.

Families without public code are not represented (history-based retrieval) or are represented by a generic reimplementation (decomposition). The In-N-Out graph retriever \cite{r4} takes the gold target API and a missing parameter as input, so it is not a query-level retriever and is not included. We also do not run full exposure, which needs no retrieval but places the whole catalog in the prompt: in the function-calling format, the AppWorld catalog is 64,874 tokens and the MCP-Bench catalog 59,278 tokens under the DeepSeek-V3.2 tokenizer. Both fit in the model's context, so full exposure is feasible on these testbeds; its cost is paid in every agent call rather than once at retrieval, and it is a question of execution cost and tool selection rather than of retrieval completeness.

\section{Results}
\subsection{Completeness}
Table 2 reports completeness and recall on both testbeds; Appendix A gives the values by subset. Four observations stand out.

\begin{table}[htbp]
\centering\small
\caption*{\textbf{Table 2.} Completeness at k (C@k) and recall at 16 (R@16). AppWorld: 345 test queries; MCP-Bench: 103 tasks. The last column of each block is the paired bootstrap 95\% confidence interval of C@16 minus COLT's C@16.}
\begin{adjustbox}{max width=\linewidth}
\begin{tabular}{llrrrrrrrrrr}
\toprule
Family & Arm & AppWorld C@4 & C@8 & C@16 & R@16 & $\Delta$C@16 vs COLT & MCP-Bench C@4 & C@8 & C@16 & R@16 & $\Delta$C@16 vs COLT \\ \midrule
Flat & BM25 & .000 & .000 & .000 & .188 & [-.377, -.275] & .049 & .097 & .126 & .395 & [-.553, -.359] \\
Flat & bge-base & .000 & .000 & .020 & .190 & [-.354, -.258] & .194 & .233 & .272 & .676 & [-.398, -.223] \\
Flat & e5-base & .009 & .014 & .064 & .336 & [-.313, -.209] & .165 & .252 & .330 & .717 & [-.340, -.175] \\
Decomposition & LLM split + bge & .000 & .012 & .035 & .417 & [-.342, -.241] & .165 & .223 & .359 & .776 & [-.311, -.136] \\
Hierarchical & AnyTool & .107 & .212 & .301 & .757 & [-.081, .029] & .175 & .223 & .398 & .783 & [-.282, -.087] \\
Server-then-tool & MCP-Zero & .000 & .000 & .000 & .234 & [-.377, -.275] & .194 & .223 & .243 & .638 & [-.437, -.252] \\
Dependency graph & GoS & .000 & .000 & .096 & .366 & [-.281, -.180] & .214 & .223 & .243 & .536 & [-.437, -.252] \\
Set-level (trained) & COLT & .110 & .220 & .328 & .702 &  & .252 & .330 & .583 & .887 &  \\
Set-level (trained) & HYSET (simplified) & .000 & .000 & .000 & .213 & [-.377, -.275] & .019 & .019 & .039 & .148 & [-.641, -.447] \\
\bottomrule
\end{tabular}
\end{adjustbox}
\end{table}

First, completeness is low for every structure that is not trained on the testbed or guided by many LLM calls. On AppWorld, flat retrieval, decomposition, server-then-tool matching and the dependency graph reach at most .096 at k = 16. BM25 completes no AppWorld query and 12.6\% of MCP-Bench tasks; its recall at 16 is close to bge on AppWorld (.188 vs .190) and well below it on MCP-Bench (.395 vs .676). MCP-Zero's lists contain at most 12 APIs on AppWorld (3 tools per request, merged without duplicates, with a median of one request), so its completeness at 16 is capped by construction.

Second, two arms reach about a third of AppWorld queries. COLT reaches .328 and AnyTool .301 at k = 16, and their difference is not distinguishable from zero ([-.081, .029]). They split by difficulty: COLT completes .589 of test-normal and .171 of test-challenge queries, AnyTool .426 and .227.

Third, on MCP-Bench COLT leads every other arm at k = 16 (all intervals exclude zero), with .583 completeness and .887 recall. Its completeness by number of servers is .782, .467 and .167 for one, two and three servers. No other arm completes any three-server task except AnyTool (.278), and only AnyTool, decomposition, e5, bge and MCP-Zero complete any two-server task (.200 or less apart from decomposition's .233).

Fourth, recall and completeness diverge. Decomposition raises AppWorld recall at 16 from .190 (bge) to .417 but completeness only from .020 to .035. On MCP-Bench, AnyTool and decomposition reach recall at 16 of .783 and .776 against COLT's .887, but completeness of only .398 and .359 against .583.

The HYSET arm is the simplified variant described in Section 4.3 (no execution reward, larger maximum set size), so its numbers do not represent the full method. On AppWorld it completes no query at k = 16 (recall .213); on MCP-Bench its completeness at 16 is .039 and its recall .148.

The dependency-graph arm is sensitive to how its graph is built. Without the LLM (deterministic edges only), GoS reaches similar AppWorld completeness at 16 (.119 vs .096; the paired interval of the difference includes zero) but higher recall (.474 vs .366; the LLM build is lower by between .085 and .130). On MCP-Bench, where deterministic edges cannot form, the LLM build raises completeness at 16 from .194 to .243.

\subsection{Retrieval-Time Cost}
Table 3 lists what each arm spends per query once it is built, and what it requires beforehand.

\begin{table}[htbp]
\centering\small
\caption*{\textbf{Table 3.} Retrieval-time cost per query (median over queries; DeepSeek-V3.2 through a commercial API, so latency includes network and queueing) and prerequisites. AW = AppWorld, MB = MCP-Bench.}
\begin{adjustbox}{max width=\linewidth}
\begin{tabular}{lrrrl}
\toprule
Arm & LLM calls (AW / MB) & Tokens (AW / MB) & LLM latency, s (AW / MB) & Requires before use \\ \midrule
BM25, bge, e5 & 0 / 0 & 0 / 0 & 0 / 0 & nothing \\
Decomposition & 1 / 1 & 221 / 580 & 5.7 / 15.7 & nothing \\
MCP-Zero & 1 / 1 & 333 / 697 & 4.4 / 12.3 & nothing \\
AnyTool & 31 / 38 & 99,739 / 90,479 & 307 / 311 & app or server grouping \\
GoS & 0 / 0 & 0 / 0 & 0 / 0 & graph built with LLM calls \\
COLT & 0 / 0 & 0 / 0 & 0 / 0 & labelled queries on the same pool \\
HYSET (simplified) & 0 / 0 & 0 / 0 & 0 / 0 & labelled queries on the same pool \\
\bottomrule
\end{tabular}
\end{adjustbox}
\end{table}

The two arms that complete the most queries spend in different places. COLT spends nothing at query time beyond encoding but requires labelled queries over the same tool pool: 90 on AppWorld and 75 per fold on MCP-Bench. AnyTool needs no training data but spends a median of 31 LLM calls and about 100 thousand tokens per AppWorld query, about five minutes of wall-clock time. The one-call arms (decomposition and MCP-Zero) add a few hundred tokens and several seconds per query and do not raise AppWorld completeness above .035.

Figure 3 places each arm by completeness at 16 and the LLM tokens it spends per query at retrieval time.

\begin{figure}[htbp]
\centering
\includegraphics[width=\linewidth]{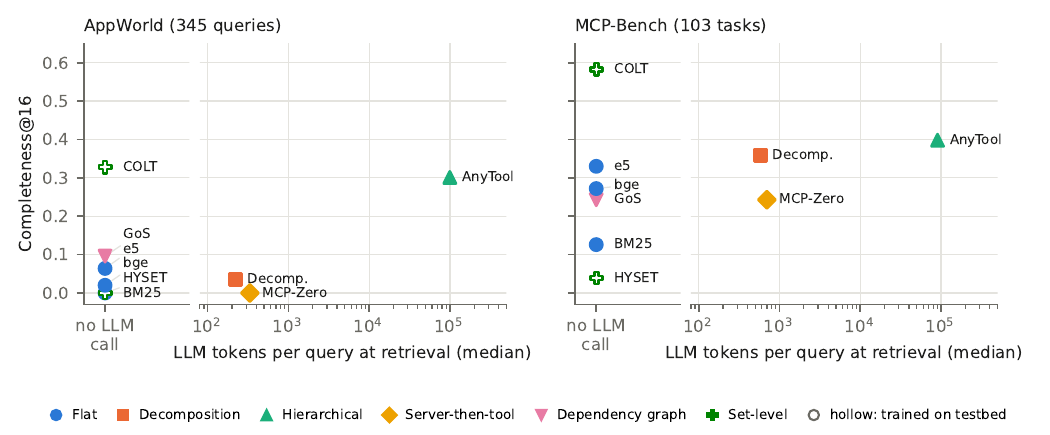}
\caption*{\textbf{Figure 3.} Completeness at 16 against median LLM tokens per query at retrieval time. Arms without a query-time LLM call appear in the left strip of each panel; GoS calls the LLM only when building its graph. Hollow markers are trained on the testbed's labelled queries.}
\end{figure}

\subsection{Why Retrieval Is Incomplete}
For each incomplete query we classify every missing required API at k = 16 by depth (ranked 17 to 50 in the arm's list, or absent from it) and, on AppWorld, by role in the query's dependency graph: a target is an API that no other required API consumes; a prerequisite produces an input for another required API, and is unrecognized if at least one of its consumers is in the top 16 and part of a cascade otherwise. A query's primary cause is the first that applies in the order target, unrecognized prerequisite, cascade, alternative only. In-N-Out lists alternative producers for 48 of 3,801 parameters, and at most 6 missing APIs per arm have a retrieved alternative, so strict scoring does not penalize arms that choose another producer.

\begin{table}[htbp]
\centering\small
\caption*{\textbf{Table 4.} Failure decomposition at k = 16 on AppWorld. Primary cause counts incomplete queries; beyond k is the share of missing APIs ranked 17 to 50.}
\begin{adjustbox}{max width=\linewidth}
\begin{tabular}{lrrrr}
\toprule
Arm & Incomplete / 345 & Primary cause: target & Primary cause: unrecognized prerequisite & Missing APIs beyond k \\ \midrule
COLT & 232 & 163 & 69 & 58\% (404 / 698) \\
AnyTool & 241 & 79 & 162 & 11\% (61 / 538) \\
Decomposition & 333 & 190 & 143 & 30\% (366 / 1,226) \\
GoS & 312 & 250 & 62 & 8\% (110 / 1,374) \\
e5 & 323 & 283 & 40 & 33\% (472 / 1,441) \\
bge & 338 & 284 & 54 & 24\% (405 / 1,686) \\
BM25 & 345 & 282 & 63 & 21\% (351 / 1,639) \\
HYSET (simplified) & 345 & 340 & 5 & 25\% (390 / 1,559) \\
MCP-Zero & 345 & 236 & 109 & 0\% (0 / 1,549) \\
\bottomrule
\end{tabular}
\end{adjustbox}
\end{table}

Three failure patterns emerge. COLT ranks part of the required set just below the cutoff: 58\% of its missing APIs on AppWorld and 90\% on MCP-Bench are in its top 50. Its failures are therefore ones of ranking depth, which a larger budget or a reranker could address. AnyTool's lists are short (258 of 345 pools hold fewer than 16 APIs) and its dominant failure is the unrecognized prerequisite: it reaches the app and API that the request names and stops without the APIs that supply their inputs; 317 of its 538 missing APIs are of this kind. Flat and graph arms and the simplified HYSET mostly miss the target itself. MCP-Zero's lists end by rank 12, so none of its missing APIs fall in ranks 17 to 50. For decomposition, 602 of its 1,226 missing APIs are unrecognized prerequisites. On MCP-Bench, which has no dependency labels, the depth pattern is the same: missing tools ranked 17 to 50 make up 90\% for COLT, 66\% for decomposition, 36\% for AnyTool and 3\% or less for GoS and MCP-Zero, whose lists often end before rank 16.

\subsection{Dependency Structure of the Need}
We split AppWorld queries by the shape of their required-API dependency graph. On the 258 single-chain queries, AnyTool and COLT are tied (C@16 .376 and .372; difference [-.062, .066]), ahead of the deterministic GoS (.159), GoS (.128) and all other arms (.085 or less). On the 87 mixed queries, completeness is lower for every arm: COLT reaches .195, AnyTool .080 (difference [-.218, -.011]) and every other arm 0. Mixed queries, however, need more APIs (median 7 vs 4), and no mixed query needs fewer than 5. We therefore compared the two shapes at equal required-set size, over sizes 5 to 12 where both occur (87 mixed and 120 chain queries), weighting the per-size differences by the number of mixed queries (Table A5). At equal size, the size-standardized difference in C@16 (mixed minus chain) is -.065 for AnyTool (95\% bootstrap interval [-.153, .019]) and +.132 for COLT ([.049, .217]); at these sizes, decomposition completes 2 of the 207 queries and the flat, graph, MCP-Zero and HYSET arms complete none. At equal size, COLT completes mixed queries more often, and AnyTool's interval runs from a loss of .153 to a gain of .019, so this split detects no loss of completeness from a mixed shape.

\subsection{Downstream Check}
To test whether completeness matters for task success, we ran the official AppWorld simplified ReAct code agent with its prompt and DeepSeek-V3.2 configuration (temperature 0, at most 50 steps) on 60 test tasks sampled at random within each split (20 test-normal, 40 test-challenge), under two exposures: the gold required APIs (A) and COLT's top 16 (C). In both, only the exposed APIs (plus the supervisor) can be documented and called, and calls to others return AppWorld's own errors.

Exposure A succeeded on 46 of 60 tasks and C on 23 (25 tasks succeeded only under A and 2 only under C; exact McNemar p = 5.6 $\times$ 10\^{}-6). The difference comes almost entirely from incomplete retrieval. On the 21 tasks where COLT's top 16 was complete, C succeeded on all 21 and A on 20. On the 39 incomplete tasks, C succeeded on 2 and A on 26. Under C the agent took more steps (median 32 vs 19), hit an unexposed-API error in 30 episodes (vs 8 under A), and used about twice the prompt tokens (15.1 million vs 7.4 million). The comparison between complete and incomplete tasks is observational, since complete tasks may be easier; but on the same incomplete tasks the gold exposure succeeds two thirds of the time, so the gap is not explained by task difficulty alone.

\section{Discussion}
Single-tool evaluation misranks retrievers for multi-tool needs. On ToolRet, the retriever that is best on single-target queries is overtaken on multi-target ones by a retriever trained on tool data and by decomposition. Together with the crossover reported by Zhang et al. \cite{r10}, this suggests that aggregate scores dominated by single-target queries should not be used to choose a retriever for agents whose requests need several tools.

Completeness on multi-tool requests is bought with either labelled data or LLM calls. On our testbeds, none of the compared structures that runs zero-shot with at most one LLM call completes more than 10\% of AppWorld queries. The two that do better spend different resources: in-domain labelled queries for COLT and tens of LLM calls per query for AnyTool. Which to prefer depends on whether the tool pool comes with representative labelled requests and on how much query-time latency the application tolerates. Decomposition with one call raises recall but rarely completes a set.

The failure patterns point to different repairs. COLT's misses are mostly at ranks 17 to 50, so a larger budget or a set-aware reranker over its top 50 could recover many of them. AnyTool's misses are prerequisites of APIs it already found, which suggests completing a found set along producer-consumer dependencies after navigation. Flat and graph arms miss targets, so their problem lies at the first retrieval step. Graph-based expansion helps only when edges exist: MCP tools in MCP-Bench declare no output schemas, so Graph-of-Skills cannot form its deterministic edges there, and its LLM-built edges reduce recall on AppWorld relative to the deterministic graph.

Completeness is a useful proxy for downstream success in this setting. Complete sets almost always led to success, and incomplete sets almost never did, while the gold set succeeded on most of the same incomplete tasks.

\section{Limitations}
All LLM steps use DeepSeek-V3.2, so the completeness and call counts of the LLM-dependent structures, and with them the trade-off in Section 6 between labelled data and LLM calls, are measured for this one model. Hierarchical discovery depends on model capability more than flat exposure does: in GatewayBench \cite{r18}, replacing GPT-5.4 with GPT-5.1 lowered the task success of hierarchical gateways by 37.6 points on average, against 11.9 points for flattened ones. AnyTool's position relative to COLT, which calls no LLM at query time, is therefore tied to the capability of the navigating model.

The dense arms, and the decomposition, server-then-tool and graph arms built on them, use base-size embedding models (bge-base and e5-base, the pair ToolRet also evaluates). Larger general-purpose and tool-specific embedding models score higher on ToolRet and its re-annotation \cite{r8,r16}, so the 10\% ceiling in Section 6 describes these structures with encoders of this size, and evaluating them with stronger encoders is future work.

\section{Conclusion}
Retrieving a complete tool set decides whether a multi-tool request succeeds in our downstream check: complete sets succeeded 21 of 21 times and incomplete ones 2 of 39 times, and benchmarks dominated by single-tool queries hide how rarely retrievers achieve it. Across six retrieval structures on two multi-tool testbeds, the choice follows from cost: when labelled requests exist for the tool pool, trained set-level retrieval, as represented by COLT, gives the most complete sets at no query-time LLM cost; without them, only LLM-navigated hierarchical search, as represented by AnyTool, reaches a comparable level, at 31 to 38 LLM calls and about five minutes per query; and with at most one LLM call per query, none of the flat, decomposition, server-then-tool and dependency-graph structures completes a tenth of AppWorld requests. The two leading structures fail in different ways, which motivates different repairs: most tools the set-level retriever misses are already in its top 50, within reach of a reranker, and the hierarchy misses prerequisites of the APIs it finds, which dependency completion targets.

\clearpage
\appendix
\section{Full Results}
Tables A1 and A2 give completeness at k = 4, 8 and 16 and recall at 16 for every arm by subset of the two testbeds, including the deterministic Graph-of-Skills ablation. Tables A3 and A4 give completeness and recall at 10 on ToolRet by number of targets, with and without the three ToolBench-derived pools. Table A5 gives the per-size values behind the size-controlled comparison in Section 5.4; several cells contain fewer than ten queries.

\begin{table}[htbp]
\centering\small
\caption*{\textbf{Table A1.} AppWorld: completeness at k and recall at 16 by subset (normal n=129, challenge n=216, all n=345).}
\begin{adjustbox}{max width=\linewidth}
\begin{tabular}{lrrrrrrrrrrrr}
\toprule
Arm & normal C@4 & C@8 & C@16 & R@16 & challenge C@4 & C@8 & C@16 & R@16 & all C@4 & C@8 & C@16 & R@16 \\ \midrule
BM25 & .000 & .000 & .000 & .254 & .000 & .000 & .000 & .148 & .000 & .000 & .000 & .188 \\
bge-base & .000 & .000 & .039 & .279 & .000 & .000 & .009 & .137 & .000 & .000 & .020 & .190 \\
e5-base & .023 & .039 & .093 & .383 & .000 & .000 & .046 & .308 & .009 & .014 & .064 & .336 \\
Decomposition & .000 & .008 & .062 & .548 & .000 & .014 & .019 & .338 & .000 & .012 & .035 & .417 \\
AnyTool & .155 & .333 & .426 & .838 & .079 & .139 & .227 & .708 & .107 & .212 & .301 & .757 \\
MCP-Zero & .000 & .000 & .000 & .304 & .000 & .000 & .000 & .193 & .000 & .000 & .000 & .234 \\
GoS & .000 & .000 & .116 & .371 & .000 & .000 & .083 & .364 & .000 & .000 & .096 & .366 \\
GoS, deterministic (ablation) & .000 & .023 & .124 & .461 & .000 & .005 & .116 & .481 & .000 & .012 & .119 & .474 \\
COLT & .271 & .488 & .589 & .866 & .014 & .060 & .171 & .603 & .110 & .220 & .328 & .702 \\
HYSET (simplified) & .000 & .000 & .000 & .349 & .000 & .000 & .000 & .132 & .000 & .000 & .000 & .213 \\
\bottomrule
\end{tabular}
\end{adjustbox}
\end{table}

\begin{table}[htbp]
\centering\small
\caption*{\textbf{Table A2.} MCP-Bench: completeness at k and recall at 16 by subset (1 server n=55, 2 servers n=30, 3 servers n=18).}
\begin{adjustbox}{max width=\linewidth}
\begin{tabular}{lrrrrrrrrrrrr}
\toprule
Arm & 1 server C@4 & C@8 & C@16 & R@16 & 2 servers C@4 & C@8 & C@16 & R@16 & 3 servers C@4 & C@8 & C@16 & R@16 \\ \midrule
BM25 & .091 & .182 & .236 & .469 & .000 & .000 & .000 & .300 & .000 & .000 & .000 & .327 \\
bge-base & .364 & .436 & .473 & .785 & .000 & .000 & .067 & .621 & .000 & .000 & .000 & .432 \\
e5-base & .309 & .473 & .564 & .814 & .000 & .000 & .100 & .661 & .000 & .000 & .000 & .517 \\
Decomposition & .309 & .418 & .545 & .860 & .000 & .000 & .233 & .720 & .000 & .000 & .000 & .609 \\
AnyTool & .327 & .400 & .545 & .794 & .000 & .033 & .200 & .780 & .000 & .000 & .278 & .755 \\
MCP-Zero & .364 & .400 & .418 & .676 & .000 & .033 & .067 & .608 & .000 & .000 & .000 & .571 \\
GoS & .400 & .418 & .455 & .682 & .000 & .000 & .000 & .404 & .000 & .000 & .000 & .313 \\
GoS, deterministic (ablation) & .345 & .364 & .364 & .597 & .000 & .000 & .000 & .332 & .000 & .000 & .000 & .289 \\
COLT & .473 & .582 & .782 & .945 & .000 & .067 & .467 & .843 & .000 & .000 & .167 & .785 \\
HYSET (simplified) & .036 & .036 & .073 & .177 & .000 & .000 & .000 & .115 & .000 & .000 & .000 & .112 \\
\bottomrule
\end{tabular}
\end{adjustbox}
\end{table}

\begin{table}[htbp]
\centering\small
\caption*{\textbf{Table A3.} ToolRet, all 35 pools: completeness at 10 and recall at 10 by number of targets (n = 4,402 / 1,937 / 1,089 / 533 / 7,961); last column: paired bootstrap 95\% CI of the C@10 difference from bge-base on single-target queries.}
\begin{adjustbox}{max width=\linewidth}
\begin{tabular}{lrrrrrrrrrrr}
\toprule
Retriever & 1: C@10 & R@10 & 2: C@10 & R@10 & 3: C@10 & R@10 & 4+: C@10 & R@10 & All: C@10 & R@10 & Single-target $\Delta$C@10 vs bge \\ \midrule
bge-base & .672 & .672 & .426 & .612 & .169 & .468 & .129 & .459 & .507 & .615 &  \\
e5-base & .635 & .635 & .405 & .591 & .174 & .458 & .118 & .475 & .481 & .589 & [-.047, -.026] \\
COLT stage 1 & .642 & .642 & .525 & .701 & .234 & .522 & .129 & .482 & .523 & .629 & [-.041, -.018] \\
Decomposition & .664 & .664 & .450 & .633 & .233 & .527 & .169 & .588 & .520 & .633 & [-.018, .002] \\
GoS (deterministic) & .583 & .583 & .255 & .460 & .092 & .351 & .079 & .374 & .402 & .508 & [-.100, -.077] \\
\bottomrule
\end{tabular}
\end{adjustbox}
\end{table}

\begin{table}[htbp]
\centering\small
\caption*{\textbf{Table A4.} ToolRet, without the ToolBench-derived pools: completeness at 10 and recall at 10 by number of targets (n = 4,142 / 1,155 / 609 / 444 / 6,350); last column: paired bootstrap 95\% CI of the C@10 difference from bge-base on single-target queries.}
\begin{adjustbox}{max width=\linewidth}
\begin{tabular}{lrrrrrrrrrrr}
\toprule
Retriever & 1: C@10 & R@10 & 2: C@10 & R@10 & 3: C@10 & R@10 & 4+: C@10 & R@10 & All: C@10 & R@10 & Single-target $\Delta$C@10 vs bge \\ \midrule
bge-base & .684 & .684 & .448 & .637 & .207 & .505 & .128 & .484 & .557 & .644 &  \\
e5-base & .649 & .649 & .410 & .598 & .197 & .493 & .119 & .496 & .525 & .614 & [-.046, -.025] \\
COLT stage 1 & .650 & .650 & .494 & .687 & .241 & .520 & .110 & .460 & .545 & .631 & [-.047, -.023] \\
Decomposition & .675 & .675 & .478 & .661 & .296 & .580 & .180 & .625 & .568 & .660 & [-.020, .001] \\
GoS (deterministic) & .599 & .599 & .310 & .521 & .154 & .440 & .092 & .415 & .468 & .557 & [-.098, -.073] \\
\bottomrule
\end{tabular}
\end{adjustbox}
\end{table}

\begin{table}[htbp]
\centering\small
\caption*{\textbf{Table A5.} AppWorld chain and mixed queries at equal required-set size: C@16 (chain / mixed) for the two arms that complete any of these queries.}
\begin{adjustbox}{max width=\linewidth}
\begin{tabular}{rrrr}
\toprule
Required APIs & n chain / mixed & COLT & AnyTool \\ \midrule
5 & 12 / 9 & .250 / .222 & .333 / .222 \\
6 & 30 / 30 & .000 / .400 & .200 / .033 \\
7 & 33 / 12 & .273 / .250 & .212 / .250 \\
8 & 21 / 3 & .000 / .000 & .048 / .000 \\
9 & 6 / 18 & .000 / .000 & .000 / .000 \\
10 & 9 / 9 & .000 / .000 & .000 / .111 \\
11 & 3 / 3 & .000 / .000 & .333 / .000 \\
12 & 6 / 3 & .000 / .000 & .000 / .000 \\
\bottomrule
\end{tabular}
\end{adjustbox}
\end{table}

\section{Implementation Details}
Shared components. BM25 (k1 = 1.5, b = 0.75) indexes the same tool text as the dense arms, lowercased and split into alphanumeric tokens. bge-base-en-v1.5 encodes tools and queries for the flat bge, decomposition, MCP-Zero and Graph-of-Skills arms, with the bge query instruction ``Represent this sentence for searching relevant passages: ''. e5-base-v2 uses its ``query: '' and ``passage: '' prefixes. Every LLM step uses DeepSeek-V3.2 through the SiliconFlow API at temperature 0; all calls are cached, so reported token counts and latencies are those of the original calls. Rankings are truncated at 50.

Decomposition prompt. The single LLM call per query uses the following prompt; the response is parsed as a JSON array, and the query itself is used when parsing fails.

\begin{lstlisting}
Split the user request below into the minimal list of independent sub-requests, one per distinct action or piece of information the request needs (include implicit steps such as looking up an item before acting on it). Each sub-request must be a short self-contained sentence. If the request is a single action, return it unchanged as a one-item list.
Return only a JSON array of strings.

Request: {query}
\end{lstlisting}

Each sub-request retrieves its top 50 tools with bge, and the per-sub-request lists are merged round-robin with duplicates removed.

MCP-Zero. We use the request prompt and matcher logic of the released code (commit fd666c4): servers are ranked by the similarity between the requested server and the server description, the top 5 are kept, and tools in them are scored by (s$\cdot$t)$\cdot$max(s, t) with the top 3 kept per request. Three changes are made. First, bge replaces the commercial embedding model, so the embedding model is held fixed across arms. Second, neither testbed provides server READMEs, so the server text is the placeholder that AnyTool also receives (``The <app> app.'' or ``The <server> MCP server.''). Third, MCP-Zero requests tools one at a time during execution, which a retrieval-only run cannot do, so one sentence added to the prompt asks for one request block per needed tool in a single call; the per-request lists are merged round-robin.

AnyTool. The official agent code is used unchanged except for three adapters: apps or servers serve as both the category and the tool level of its hierarchy, the GPT calls are routed to DeepSeek-V3.2 through the tools API, and the API-list function schema is changed from a null type to an array of objects. The search stops under the official rules (its own solvability check, at most 64 APIs, 5 leaf tools per node and a 200,000-token cap), and the API pool at that point is the arm's output; the downstream solver and self-reflection are not run.

Graph-of-Skills. Each tool is converted to a skill document listing its name, description, inputs and outputs; MCP tools publish no output schema, so their output list is empty and no deterministic edge forms on MCP-Bench. The reported arm uses the default build, with full-markdown indexing, LLM semantic completion of input-output edges and LLM-validated relation edges; query rewriting is off, its default. The deterministic ablation disables the LLM steps and keeps everything else, including full-markdown indexing.

COLT. On ToolBench G2 we ran the authors' configuration, changed only for speed, and obtained R@5 85.26, N@5 82.78 and C@5 70.85 at the best epoch (12 of 34, early stopping), against 85.03, 82.54 and 70.10 in the COLT paper. In the released data the tuning and test files are identical, so this epoch selection follows the authors' protocol and uses the test set. For AppWorld and MCP-Bench, the first-stage training loop is reimplemented in plain PyTorch with the same recipe because the authors' training wrapper is incompatible with current library versions; the batch size and number of negatives are adapted to the small training sets (batch size 32 on both; 20 negatives on AppWorld, whose sets reach 13 tools, and 40 on MCP-Bench, whose sets reach 20), and checkpoints are selected on the development split at k = 10. AppWorld models are trained without the supervisor APIs.

HYSET. The simplified variant uses a frozen bge-base encoder, sets the execution-reward weight to zero, and raises the maximum set size from 5 to the largest required set (13 on AppWorld, 20 on MCP-Bench); other hyperparameters follow the paper's defaults. Rankings list the predicted set followed by the remaining tools in singleton-score order. The released code includes the judge step of the execution reward but not the agent rollouts it scores.

MCP-Bench required sets. MCP-Bench provides no tool-set labels, so we derive them. A tool belongs to a task's required set if it belongs to one of the task's servers and is named in the task's detailed description or dependency analysis. Names containing an underscore or hyphen match as whole tokens; plain-word names (such as ``sum'' or ``fetch'') match only in call context: ``Server:name'', ``name('', a backquoted name, or after a calling verb such as ``call'', ``use'', ``invoke'' or ``via''. An underscore-joined token that is the prefix of exactly one tool name on the task's servers also counts. One task (osint\_intelligence\_000) names its tools only by paraphrase and is excluded, leaving 103. MCP-Bench folds for COLT and HYSET are stratified by the number of servers (seed 42), with 10\% of each training fold held out for checkpoint selection.

Downstream exposure. The agent is a port of the official simplified ReAct code agent with its prompt file and DeepSeek-V3.2 configuration (temperature 0, seed 100, at most 50 steps, no prompt or output trimming), run on AppWorld 0.1.3, the version that matches the In-N-Out labels. The supervisor APIs are always exposed. Hidden APIs are deleted from the environment, the API documentation tools list only exposed apps and APIs and return AppWorld's own error messages for others, the prompt's app list is filtered the same way, and the raw request object is removed from the code interpreter so the filter cannot be bypassed. A prompt longer than the endpoint's 163,840-token limit counts as a failure of the episode.

\end{document}